\documentclass[iop]{emulateapj}

\usepackage{amsmath}
\usepackage{amssymb}
\usepackage{graphicx}
\usepackage{natbib}
\usepackage[backref,breaklinks,colorlinks,citecolor=blue]{hyperref}
\usepackage{sidecap}
\usepackage{comment}

\newcommand{\WMAP}{\textsl{WMAP}}

\newcommand{\wmap}{{\WMAP}}
\newcommand{\Planck}{{\textsl{Planck}}}
\newcommand{\planck}{{\textsl{Planck}}}
\newcommand{\lcdm}{\ensuremath{\Lambda}CDM}

\renewcommand{\ell}{\ensuremath{l}}
\newcommand{\be}{\begin{equation}}
\newcommand{\ee}{\end{equation}}

\newcommand{\beq}{\begin{equation}}
\newcommand{\eeq}{\end{equation}}
\newcommand{\beqa}{\begin{eqnarray}}
\newcommand{\eeqa}{\end{eqnarray}}

\def\ba{\begin{eqnarray}}
\def\ea{\end{eqnarray}}

\newcommand{\barr}{\begin{array}}
\newcommand{\earr}{\end{array}}

\usepackage[flushleft]{threeparttable}
\usepackage{array}
\newcolumntype{C}[1]{>{\centering\let\newline\\\arraybackslash\hspace{0pt}}m{#1}}

\shorttitle{Effect of Template Uncertainties on $\tau$}
\shortauthors{Weiland et al.}
\slugcomment{Accepted for publication by the Astrophysical Journal}

\begin{document}

\title{Effect of Template Uncertainties on the \wmap\ and \planck\ Measures of the Optical Depth due to Reionization}

\author{J. L. Weiland, K. Osumi, G. E. Addison, C. L. Bennett, D. J. Watts}
\affil{Department of Physics and Astronomy, Johns Hopkins University, 3400 N. Charles St, Baltimore, MD 21218, USA}
\email{jweilan2@jhu.edu}
\and
\author{M. Halpern, G. Hinshaw}
\affil{Department of Physics and Astronomy, University of British Columbia, Vancouver, BC, V6T 1Z1, Canada }

\begin{abstract}
The reionization optical depth is the most   
poorly determined of the six \lcdm\ parameters fit to CMB anisotropy  
data.  Instrumental noise and systematics have prevented uncertainties from reaching  
their cosmic variance limit. 
At present, the datasets providing the most statistical constraining power are  
the \wmap, \planck\ LFI, and
\planck\ HFI full-sky polarization maps.  
As the reprocessed HFI data with reduced systematics
are not yet publicly unavailable,
we examine determinations of $\tau$ using 9-year \wmap\  and 2015 \planck\ LFI  data,   
with an emphasis on characterizing potential systematic bias resulting from foreground template and masking choices. 
We find evidence for a low-level systematic in the LFI polarization data with a   
roughly common-mode morphology across the LFI frequencies and a spectrum   
consistent with leakage of intensity signal into the polarization channels.    
We demonstrate significant bias in the optical depth derived when using the LFI 30 GHz map as   
a template to clean synchrotron from \wmap\ data, and  
recommend against the use of the 2015 LFI 30 GHz polarization data as a foreground template  
for non-LFI datasets.  We find an inconsistency between versions of the 2015 polarized 353~GHz  
dust templates reconstructed from the \planck\ likelihood and those from delivered maps, which  
can affect $\tau$ at the $1\sigma$ level.  
The spread in $\tau$ values over the ensemble of data combinations we study
suggests that systematic uncertainties still contribute significantly to the current uncertainty in $\tau$,
but all values are consistent with the range of $\tau = 0.07 \pm 0.02$.
\end{abstract}


\section{Introduction}

The optical depth due to reionization, $\tau$, is one of the six defining parameters of the standard
\lcdm\ cosmological model.
The value of $\tau$ provides a measure of the
line-of-sight free electron opacity to CMB radiation, and is relevant to how and when the first stars and galaxies formed.  
Additionally, $\tau$ shares mutual degeneracies
with other \lcdm\ parameters \citep{zaldarriaga/spergel/seljak:1997}, and smaller $\tau$ uncertainties translate
to more tightly constrained determinations of the curvature power $A_s$, the amplitude of linear matter fluctuations $\sigma_8$, and indirectly the Hubble
constant $H_0$.  
Uncertainty in $\tau$ will limit cosmological detection of neutrino mass unless current measurements can be significantly improved 
upon \citep[e.g.][]{allison/etal:2015, calabrese/etal:2017, boyle/komatsu:2017, watts/etal:2018}.

Although substantial progress has been made in the last few years, $\tau$ remains
the least well-determined of the six \lcdm\ parameters.
Lower limits on $\tau$ may be derived from the lack of the Gunn-Peterson trough \citep{gunn/peterson:1965} in the spectra of distant 
quasars and galaxies at redshifts z $\approx 6$ \citep[e.g.][]{pryke/etal:2002, fan/etal:2005, planck/intermediate/46:2016}.
However, CMB data still provide the best means of measuring $\tau$.
Specifically, the reionization bump located at power spectrum 
multipole moments $\ell < 15$ in the E-mode of the CMB linear polarization is the most direct 
means for a detection, although this measurement is observationally challenging.  
Cosmic variance places a lower limit on uncertainties that may be quoted
for these low-\ell\  polarization determinations:  
$\sigma(\tau)/\tau = 2.5\%$ for a full sky analysis, or roughly $\sigma(\tau) = 0.002$ \citep{calabrese/etal:2017, watts/etal:2018}.
The cosmic variance limit has yet to be achieved.
Low multipole polarization observations
are often combined with temperature and other datasets in a full likelihood to maximize constraining power on parameters
\citep[e.g.][]{hinshaw/etal:2013, planck/11:2015}.

In Table~\ref{tab:tau_pub_tab}, we show a sample of the range of determinations of $\tau$ over the
last few years, as derived from \wmap\ and \planck\ CMB observations, using a variety of data combinations,
analyzed sky regions, and foreground removal options.  This is not a comprehensive list, but it does
encompass a representative range in values.  We will refer to some of the table entries throughout
this discussion.

\begin{deluxetable*}{lll}
\tablecaption{Recently published values for $\tau$ \label{tab:tau_pub_tab}}
\tablewidth{5in}
\tablehead{
\colhead{CMB Data} & 
\colhead{$\tau$} &
\colhead{Reference}
}
\startdata
\wmap\  TT+poln+BAO+eCMB\tablenotemark{a}  &   $0.089 \pm 0.014$    &  \citet{bennett/etal:2013} \\
\planck\ TT + LFI poln\tablenotemark{b}               &   $0.078 \pm 0.019$    &  \citet{planck/13:2015}  \\
\wmap\  + LFI low-\ell\tablenotemark{c}               &   $0.071 \pm 0.012$    &  \citet{planck/11:2015} \\
\wmap\ low-\ell\tablenotemark{d}                             &   $0.067 \pm 0.013$      & \citet{planck/11:2015} \\
\planck\ LFI poln\tablenotemark{e}                       &   $0.067 \pm 0.023$    &  \citet{planck/11:2015} \\
\planck\ TT+lens+BAO\tablenotemark{f}             &   $0.067 \pm 0.016$    &  \citet{planck/intermediate/46:2016} \\
\wmap\ TT+lens+BAO\tablenotemark{g}             &   $0.066 \pm 0.020$    &  this work, Section 5 \\
\planck\  HFI 100x143 GHz\tablenotemark{h}     &   $0.058 \pm 0.012$    &  \citet{planck/intermediate/47:2016} \\
\planck\  HFI 100x143 GHz\tablenotemark{h}      &   $0.055 \pm 0.009$    &  \citet{planck/intermediate/46:2016} \\
\enddata

\tablenotetext{a}{\wmap\ TT+poln+BAO+eCMB = full \wmap\ likelihood including Baryon Acoustic Oscillations (BAO) and 
SPT and ACT data}
\tablenotetext{b}{\planck\ TT + LFI poln = temperature and low multipole polarization likelihood}
\tablenotetext{c}{\wmap\ + LFI low-\ell\ = low multipole \wmap\ and \planck\ likelihood, includes TE with polarization} 
\tablenotetext{d}{\wmap\  low-\ell\ = \planck\ reanalysis of low multipole \wmap\  likelihood, includes TE with polarization} 
\tablenotetext{e}{\planck\ LFI poln = low multipole polarization likelihood only} 
\tablenotetext{f}{\planck\ TT+lens+BAO = \planck\ temperature and weak lensing combined with BAO} 
\tablenotetext{g}{\wmap\ TT+lens+BAO = \wmap\ 9-year temperature and \planck\ weak lensing combined with recent BAO}
\tablenotetext{h}{\planck\  HFI 100x143 GHz = low multipole EE polarization data analysis using cross spectra from reprocessed 100 and 143 GHz maps, with two slightly different algorithms.}

\end{deluxetable*}

Uncertainties for $\tau$ quoted in the table represent statistical uncertainties incorporating instrumental noise 
and modeled CMB signal covariance; instrumental noise is the dominant contributor.  
In the case of $\tau$ values determined from polarization data, however, 
eliminating or at least characterizing potential systematic bias is an equally critical issue, which the 
statistical errors do not reflect. Potential sources of bias include:
\begin{itemize}
\item{ {\textit{Foreground removal systematics}}.
Polarized Galactic foreground emission outshines 
CMB polarization over the entire frequency range used for CMB observations.  Foreground analyses 
to date rely on observational templates to describe the spatial behavior of the foreground emission, and 
assume a spatially uniform scaling with frequency.  Even modest errors in the template and/or its 
scaling with frequency can bias the derived value of $\tau$. 
For example, \citet{planck/11:2015} note that the 2013 \wmap\ result is roughly
$1\sigma$ lower if the \planck\ HFI 353 GHz polarization map is used as a dust emission template 
in place of \wmap's template based on 
the FDS dust intensity model \citep{finkbeiner/davis/schlegel:1999} 
and starlight polarization data.}

\item{ {\textit{Choice of analysis mask}}.
The choice of sky region to be analyzed is closely related to the accuracy of foreground removal:
an analysis mask is used to exclude regions where foreground residuals or data systematics are believed to exceed some 
acceptable threshold.   
A commonly used mask construction method employs a threshold cut in polarized foreground intensity $P$;  
foreground template coefficients are often, but not always, evaluated using the same mask.  Possible sources of 
bias are similar to those described above related to foreground removal.
Potential also exists for the introduction of more subtle forms of bias; we briefly discuss one possibility in Section~3.
 }

\item{ {\textit{Instrumental systematics}}.
If not removed, instrumental systematics may introduce spurious sky signatures that may be mistaken for cosmological signal.
For example,
the 2015 \Planck\ LFI and HFI polarization data contain known instrumental systematics that are coherent across
large angular scales and produce signal artifacts at low multipoles.
These systematics include leakage of temperature signal into the polarization channels 
(including the so-called ``bandpass mismatch'' correction, or BPM) and nonlinearities in the analog-to-digital converters (ADCs) in the bolometer readouts 
that increase effective noise levels 
and introduce spurious signatures \citep{planck/intermediate/46:2016}. Estimated BPM corrections are available for LFI frequencies
and HFI 353 GHz, and have 
been applied to some of the 2015 data products, however the corrections are a major source of uncertainty in
the polarization maps \citep{planck/02:2015, planck/08:2015}.  These calibration issues rendered
the 2015 HFI polarization maps unsuitable for direct
cosmological studies \citep{planck/08:2015}, and restricted the 2015 LFI low multipole cosmological analysis to 70 GHz maps, 
after omitting survey data that failed null tests \citep{planck/11:2015}.
}
\end{itemize}

The latest \planck\ 2016 HFI $\tau$ determinations \citep{planck/intermediate/46:2016, planck/intermediate/47:2016}, based
on reprocessed data with improved calibration, 
represent the tightest constraints on $\tau$ yet. 
\citet{planck/intermediate/46:2016} document the results of extensive efforts toward characterization and minimization of HFI 
instrumental systematics. In light of the challenging systematics removal issue, however, an independent determination of 
$\tau$ is prudent.  The reprocessed HFI data are not yet public, so we looked at other datasets.
Determinations of $\tau$ that avoid the use of large angular scale polarization data may be made if the 
partial degeneracy between $A_s$ and $\tau$ present in 
the temperature (TT) power spectrum is broken using CMB lensing or other large-scale structure (LSS) information, which is sensitive 
to the amplitude of density fluctuations but not whether photons scattered since recombination.
In Table~\ref{tab:tau_pub_tab}, we show results from two such polarization-free determinations, which combine TT,  \planck\ CMB 
lensing and BAO data.
However, uncertainties are not yet competitive with the lower statistical uncertainties quoted for the
HFI results.

In this paper, we focus on determinations of $\tau$ from the \wmap\  and  2015 LFI polarization data.  
As seen in the Table~\ref{tab:tau_pub_tab} entry,
a combined \wmap+LFI dataset  can be statistically competitive with the 2016 HFI results.  
As well as providing data independent of HFI, \wmap\ and LFI  
sample a different frequency range than HFI and test the robustness of the result to foreground removal.
The \citet{planck/11:2015} \wmap+LFI analysis produces a value for $\tau$ that is roughly $1\sigma$ larger than
the HFI result, and we wish to test if this result is stable when subjected to alternative analysis choices.  
As part of a larger effort to understand uncertainties in $\tau$,
we examine shifts in $\tau$ related to different choices of foreground removal templates applied to \wmap\ and LFI data.
We describe our method of analysis in section 2, and present individual \planck\ LFI and \wmap\ results in section 3.  
In section 4 we discuss template systematics that can lead to severely compromised values for $\tau$ 
when analyzing \wmap\ and LFI data together.  We discuss our conclusions and implications for $\tau$
in section 5.

\section{Data and Methodology}

We use publicly available CMB polarization maps and noise covariance matrices in two Stokes parameters, Q and U, from the 
2015 \planck\ release \citep{planck/01:2015} and the \wmap\ 9 year data release \citep{bennett/etal:2013} in \texttt{HEALPix} format \citep{gorski/etal:2005}. 
From \planck, we use 
the 30, 70, and 353 GHz maps and from \wmap, we use maps from the K (23~GHz), Ka (33~GHz), Q (41~GHz), and V (61~GHz) bands. 

Our approach is similar to the exact pixel-based likelihood methods adopted by both the \Planck\ collaboration and the 
\WMAP\ team for low-$\ell$ analysis \citep{page/etal:2007, planck/11:2015}. 
For analysis that uses only polarization data, we introduce a Gaussian prior on the combination 
$A_s e^{-2\tau}$ and fix values for other cosmological parameters from the intensity only results in \cite{planck/13:2015}. 

The statistical properties of the CMB signal may be described as a Gaussian distributed random field on the sky using a signal covariance 
matrix in pixel space, $S$, computed from its power spectra, $C^{XX}_{\ell}$ \citep{tegmark/deoliveira-costa:2001}. 
If the data are a simple sum of CMB signal 
and Gaussian distributed instrumental noise characterized by a noise covariance matrix, $N$, the CMB likelihood function for 
this cleaned data, $\vec{m} = [Q, U]$, may be expressed as
\begin{equation}
\label{eq:LK}
\mathcal{L}(C_{\ell}) =  \frac{1}{(2\pi)^{n_p}  |M|^{1/2}} \text{exp}\left(-\frac{1}{2}\vec{m}^T M^{-1} \vec{m}\right)
\end{equation}
where $M = S + N$ is the full covariance matrix of the cleaned data and $n_p$ is the number of pixels in each of the masked Q and U maps \citep{page/etal:2007}.
We use an MCMC sampler (\texttt{emcee}, \citealt{foremanmackey/etal:2013}) to maximize the likelihood function and fit for parameters of interest and their uncertainties.

This likelihood is complicated by the need to account for foreground contamination in the raw data. In polarization, synchrotron and thermal 
dust are expected to be the dominant foregrounds at low and high frequencies, respectively. Taking advantage of the frequency dependence of 
the polarized foregrounds and CMB, a low frequency measurement may be adopted as a template for the synchrotron foreground, $\vec{T}_s$, and a high frequency 
measurement as a template for the dust, $\vec{T}_d$, to clean intermediate frequency bands where the CMB signal is relatively strong. 
We define the cleaned target map, $\vec{m}$ as
\begin{equation}
\vec{m} = \vec{m}_{\nu} - \alpha_{s} \vec{T}_{s} - \alpha_{d} \vec{T}_{d} 
\end{equation}
where $m_{\nu}$ are the observed $[Q, U]$ maps at frequency $\nu$, $\alpha_s$ scales the synchrotron template, and $\alpha_d$ scales the dust template. 
Cleaning is performed on maps in thermodynamic temperature units, and no spectral dependence constraint is placed on the template scaling parameters
$\alpha_s$ and $\alpha_d$.

The templates themselves contain CMB signal that must be accounted for.  The corresponding covariance matrix is given by
\begin{equation}
M = {(1-\alpha_{s}-\alpha_{d})^2} S + N_{\nu} + {\alpha_s}^2 N_{s} + {\alpha_d}^2 N_{d}
\end{equation}
where $N_{\nu}$, $N_{s}$ and $N_{d}$ are the noise covariance matrices for the frequency band of interest, the synchrotron template
and the dust emission template.  The equation holds for a single frequency map containing polarized CMB signal.  When
working with the three \wmap\ bands,  the target map is taken as the sum of the three bands, and the covariance matrix $N_{\nu}$
is likewise the sum of their covariance matrices, whereas the signal covariance matrix is then multiplied by ${(3-\alpha_{s}-\alpha_{d})^2}$.

In the exact case, inversion of the covariance matrix $M$ is required with each MCMC step, and the matrix dimensions corresponding to
\texttt{HEALPix} $N_{\rm{side}}=16$\footnote{\texttt{HEALPix} maps are divided into $12N_{\rm{side}}^2$ pixels, with each pixel width $\theta_{\rm{pix}}
\sim 58.6^{\circ}/N_{\rm{side}}$.  $N_{\rm{side}} = 16$ corresponds to full-sky map of 3072 pixels.  See \url{http://healpix.sourceforge.net}.}
 resolution can inhibit computational performance.  The \planck\ 2015 low-\ell\  $N_{\rm{side}}=16$ likelihood adopted approximations
that avoid repeated inversion, including pre-fitting the template coefficients, ignoring template covariance, and fixing the signal covariance
to a fiducial cosmology.

We have implemented likelihood codes for use with either $N_{\rm{side}}=16$ resolution maps (as with the \planck\ Collaboration's analysis) or 
$N_{\rm{side}}=8$ maps (as with the \wmap\ team's analysis).  As the cosmological EE signal is strongest for $\ell < 15$, both resolutions
are acceptable for determinations of $\tau$ ($\ell < 48$ vs. $\ell < 24$).  Tests using LFI data produced similar results 
regardless of which code was used. Our quoted results are from our $N_{\rm{side}}=8$ 
code implementations, as it offered greater speed of execution while recomputing and inverting the  covariance matrix $M$ at every chain point.
  We also included options for solving for the template foreground
parameters either beforehand as a separate step (``fixed'' option), or simultaneously with the cosmological parameters (``free'').
A simultaneous fit for all parameters of interest avoids making an intermediate approximation about the foreground cleaned noise
covariance matrix and the choice of a fiducial cosmology.

Low resolution ($N_{\rm{side}}=16$) LFI Stokes Q and U maps, along with the associated covariance matrices, bandpass
mismatch corrections and analysis mask, are available from the 
\Planck\ Legacy Archive
(PLA)\footnote{\url{\detokenize{https://wiki.cosmos.esa.int/planckpla2015/index.php/}} \\
\url{\detokenize{Map-making_LFI\#Noise_covariance_matrices}}}.
The LFI analysis excludes surveys 2 and 4 from use because of null-test failures and we follow this recommendation.
Similarly, we obtain the 9-year coadded \wmap\ low-resolution ($N_{\rm{side}}=16$) Stokes Q and U maps and inverse covariance matrices from 
LAMBDA\footnote{\url{https://lambda.gsfc.nasa.gov/product/map/dr5/} \\ \url{m_products.cfm}}.
The HFI 353 GHz full mission Q and U maps are available from the PLA at high resolution ($N_{\rm{side}}=2048$) and must be
degraded to low resolution. We smoothed the 353~GHz maps to $2^\circ$ full width half maximum (FWHM) resolution prior to downgrading.
In addition, we reconstructed the low-resolution ($N_{\rm{side}}=16$) dust template used in the
LFI analysis from the \planck\ likelihood code (\texttt{plc}).  As described in the Appendix, the dust template we reconstruct from the \texttt{plc} 
has large angular scale differences compared to the dust template we produce from downgrading the $N_{\rm{side}}=2048$ HFI product.
We further downgrade the $N_{\rm{side}}=16$ products to $N_{\rm{side}}=8$ using inverse noise weighting, with the exception of the smoothed high
resolution 353~GHz polarization maps, for which we use flat weighting.

{\renewcommand{\arraystretch}{1.2} 

\begin{center}
\begin{table*}
\caption{Derived values for $\tau$, EE only}
\label{tab:tau_results_tab}

\begin{threeparttable}
\begin{tabular*}{\textwidth}{ @{\extracolsep{\fill}} C{0.75cm} p{2.2cm} p{1.85cm} p{1.5cm} p{1.4cm} p{0.85cm}  C{2.0cm}   C{1.8cm}  p{1.4cm}}

\hline
\hline
\vspace{1mm}
Row &  CMB Data & Synchrotron Template & Dust Template\tnote{a}
 & fg coeffs\tnote{b} 
 & {$f_{\rm{sky}}$\tnote{c}} & {$\tau$} & {$\chi^{2}_{\rm{maxlike}}$}/dof & {Reference} \\
\hline
$1$ & LFI 70 GHz      &   LFI 30 GHz 	     &  \texttt{plc}     &  fixed  & $46\%$ &  $0.052 \pm 0.023$   &	-	  & \planck{\tnote{e}} \\
$2$ & LFI 70 GHz      &   LFI 30 GHz 	     &  \texttt{plc}     &  fixed  & $46\%$ &  $0.053 \pm 0.023$   & $728/734$	   & this work \\
$3$ & LFI 70 GHz       &	LFI 30 GHz  &  default  &  free   & $46\%$ &  $0.076 \pm 0.020$	& $823/734$	& this work \\
$4$ & \wmap\ KaQV    &   \wmap\ K      &  \texttt{plc}	&  free   & $46\%$ &  $0.045 \pm 0.019$   & $754/734$     & this work \\ 
$5$ & \wmap\ KaQV    &   \wmap\ K      &  \texttt{plc}	&  fixed\tnote{d} & $46\%$  &  $0.040 \pm 0.017$   & $752/734 $    & this work \\ 
$6$ & \wmap\ KaQV    &   \wmap\ K     &  default  &  fixed\tnote{d} & $75\%$  &  $0.058 \pm 0.017$   & $1190/1162 $	 & this work \\
$7$ & LFI 70 GHz      &	\wmap\ K	  &  \texttt{plc}      &  free	& $46\%$ &  $0.094 \pm 0.016$    & $846/734$     & this work  \\ 
$8$ & \wmap\ KaQV    &   LFI 30 GHz	  &  \texttt{plc}      &  free	& $46\%$ &  $0.37 \pm 0.02$      & $933/734$     & this work \\
\hline
\end{tabular*}

\begin{tablenotes}
\item[a]{``\texttt{plc}'' refers to the low resolution 353 GHz dust template reconstructed from the \planck\ likelihood code; ``default'' refers to the low resolution 353 GHz dust template degraded from the delivered high resolution maps.}
\item[b]{Foreground template coefficients are either free parameters in the likelihood, or fit in a separate step beforehand and are fixed parameters in the likelihood.} 
\item[c]{Fraction of sky used in analysis, corresponding to the \planck\ R1.50 analysis mask ($46\%$) and the \wmap\ P06 analysis mask ($75\%$)}
\item[d]{Foreground template coefficients were computed using a larger sky region than that used for analysis.}
\item[e]{\citet{planck/13:2015}}
\end{tablenotes}
\end{threeparttable}

\end{table*}
\end{center}

{\renewcommand{\arraystretch}{1.0} 

\section{Mission-separated fits for $\tau$}
 
Before combining LFI and \wmap\ datasets, we analyze the \wmap\ and \planck\ LFI data separately.
Either the LFI 70 GHz Stokes Q and U maps or the combined \wmap\ Ka, Q and V band polarization maps provide the CMB signal for
cosmological parameter fitting.  We further adopt the \planck\ 353 GHz map as the template for dust removal,
but as described in Section~2 and the Appendix, we examine two choices: the ``\texttt{plc}'' version, which is extracted from
the \planck\ low-$\ell$ likelihood code, and the ``default'' version, which we degrade from $N_{\rm{side}}=2048$ to lower
resolution. 
Consistent with the original LFI and \wmap\ analysis choices, we use \wmap\ K-band as the template for removing synchrotron 
foreground from the \wmap\ data, and LFI 30~GHz as the equivalent template for use in removing synchrotron from LFI 70~GHz.

We adopt the same data analysis masks as those chosen by \planck\  \citep{planck/11:2015} 
and \wmap\ \citep{page/etal:2007} for low-\ell\ polarization analysis.  These are respectively the R1.50 mask, which admits
roughly 46\% of the sky, and the P06 mask, which admits about $75\%$.
Both were selected from a suite of masks corresponding to a series of polarized foreground
intensity thresholding cuts that avoid regions of bright Galactic emission,
but the details of the exact mask selection algorithm are different between the two missions.
The \wmap\ diffuse P06 mask was chosen, ``as the best compromise between maximizing usable 
sky area while minimizing foreground contamination'' \citep{page/etal:2007}.
\citet{planck/11:2015} select an optimum mask based 
on $\chi^2$ minimization of foreground templates fit to the observed LFI 70 GHz data as a function of successively increasing sky area.
Use of this technique comes with caveats not discussed in \citet{planck/11:2015}.
Minimizing the variance in the template-cleaned map acts to maximize chance correlations between the CMB signal and
e.g. foreground templates, noise and systematics.  These  chance alignments remove power from the cleaned map, which will bias $\tau$ to lower values.
The degree of bias resulting from this effect is dependent on a number of factors, including masked
sky fraction and  spatial morphology of the foregrounds.    In the case where foreground templates perfectly match the observed foregrounds,
simulations indicate that power in a R1.50 masked cleaned map can be biased low by at much as $5-10\%$ through chance foreground alignments.
However,  chance alignments are not the only sources of error in the cleaned map.  Use of imperfect foreground templates 
in the map-cleaning process and/or the presence of remnant systematic signatures produces non-cosmological residuals 
that can produce a positive 
$\tau$ bias.  The $\chi^2$ process minimizes the combination of correlation bias (negative) and template errors (generally positive).
The situation is thus complex and requires an understanding of an experiment's dominant sources of uncertainty.  For future analyses, we suggest a study of mask-related uncertainties based on simulations.

Table~\ref{tab:tau_results_tab} shows a comparison between our results for $\tau$ (row~2) derived 
under the same template and masking conditions as those utilized in the \planck\ likelihood (row~1).
\wmap\ results for the same template and masking conditions are similar (row 4).  However,
use of the ``default'' dust template, a different analyzed sky fraction, and/or foreground coefficients
determined using a different sky fraction can produce up to $ \sim 1\sigma$ shifts
in these values (rows 3, 5, 6).

Some of these shifts may simply be statistical, while others may be from introduced bias.  The table illustrates
the need for an organized and systematic examination of results with regard to masking and template choices,
within the context of the same cosmological dataset.  We proceed toward using \wmap+LFI for that purpose. 

\begin{figure*} \begin{center}
\includegraphics[height=5.0in]{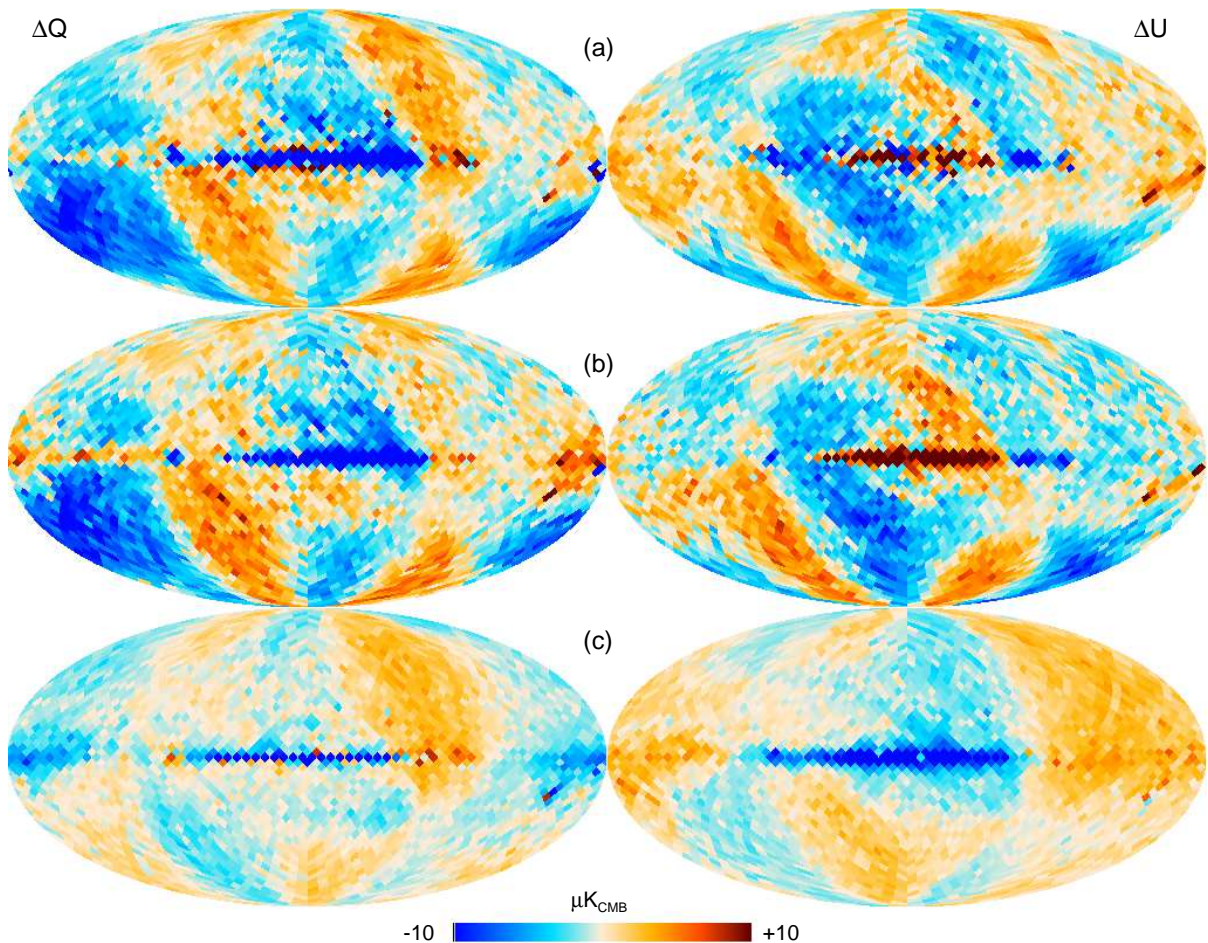}
\end{center} 
\vspace{-0.25cm}
\caption{ Differences between independent \planck\ and \wmap\ Stokes Q and U maps at similar frequencies reveal
a large angular scale signature at high Galactic latitudes that does 
not correspond to any sky signal morphology (this statement ignores the Galactic plane differences running horizontally across the
center of each map).
Such a $\Delta$Q, $\Delta$U signature has potential to influence the quality of synchrotron foreground removal through the use of 30 GHz 
and K-band templates, and thus bias the determination of $\tau$ .
\textit{(a)} The difference between the derived \planck\ \texttt{Commander} synchrotron component at
30~GHz (called $S_{30}$), and \wmap\ K-band scaled to that frequency (similar to Figure 46 of \citealt{planck/10:2015}).
\textit{(b)} Bandpass mismatch corrected LFI 30 GHz  minus \wmap\ K-band, scaled to an effective frequency of 30 GHz 
using the same $\nu^{-3,2}$ dependence as in (a).  Features similar to (a) persist, indicating that any \planck\ contribution is
not solely confined to the LFI 44 GHz band.  
\textit{(c)} $S_{30}$ minus scaled 30 GHz, or equivalently, (a)-(b), since the \wmap\ K-band contribution drops out of the difference.
This establishes the existence of large angular scale 
structure that is limited to \planck\ alone, and shares common morphological features with (a) and (b).
\label{fig:k30_qusys}
}
\end{figure*}

\section{An unaddressed systematic}

As template choices for removing synchrotron, \wmap\ K-band and LFI 30~GHz polarization maps should be almost interchangeable,
given both the similar signal-to-noise \citep{planck/25:2015} and their relative proximity in frequency.  
In a combined \wmap+LFI analysis, 
it would simplify the computation to use only one synchrotron template, and bolster confidence
in the quality of \planck\ BPM correction if results using K-band were the same as those using the LFI 30~GHz data. 

We next perform foreground cleaning of \wmap\ 
KaQV data using the \planck\ 30~GHz synchrotron template, and conversely, clean LFI 70~GHz using the \wmap\ K-band synchrotron template,
holding all other analysis assumptions constant. 
The last two rows of
Table~\ref{tab:tau_results_tab} show the resultant values for $\tau$ when using these alternate choices of the synchrotron template.
In both cases, $\chi^2$ values are significantly increased, as are the derived $\tau$ values.
The 70~GHz result returns $\tau = 0.094 \pm 0.016$, in tension with the HFI results listed
in Table~\ref{tab:tau_pub_tab}. 
\wmap\ KaQV cleaned using the LFI 30~GHz synchrotron template returned $\tau$ near $0.37$, a result that may be 
rejected as unrealistic based on existing CMB temperature observations 
(see e.g. the TT+lens+BAO entries in Table~1).
At best, neither result argues for a well-understood and
consistent representation of the polarized synchrotron template, so we needed to understand these results before proceeding further.

A solution to this problem lies with a large angular scale residual signature in Q and U that becomes apparent only when
comparing \wmap\ and LFI polarization maps.  
\citet{planck/10:2015} note the existence of large angular scale signatures unrelated to sky signal when differencing the 
empirically modeled \texttt{Commander} synchrotron component against \wmap\ K-band observations.
The difference between
the \texttt{Commander} synchrotron Q, U maps at 30~GHz ($S_{30}$) and \wmap\ K-band Q, U is performed after scaling the K-band maps to 30~GHz using a 
$\nu^{-3.2}$ power law dependence. The scaling is chosen to null the synchrotron foreground using a representative spectral index at
high Galactic latitudes.  Since synchrotron is the dominant K-band foreground emission component, the difference should primarily show 
white noise, as the polarized CMB signal is completely subdominant to instrument noise for these maps.
The Q and U components of an $S_{30} - 0.39$K difference map are shown
in the top row of Figure~\ref{fig:k30_qusys}.  The difference is dominated by a residual 
large angular scale pattern at high Galactic latitudes with a peak amplitude near $\pm 10~\mu$K, far in excess
of the expected cosmological signal.
\citet{planck/10:2015} note ``a morphology clearly associated with the scanning strategy of either \planck\ or \wmap'', and attribute the most 
likely origin of this pattern to \planck\ 44~GHz (because the 44~GHz data failed null tests at low
multipoles) and to \wmap\ low multipole ``poorly measured modes,'' which are modes of enhanced noise variance related to the \wmap\ scan
strategy \citep{page/etal:2007, jarosik/etal:2011, bennett/etal:2013}. 
In a different analysis, \citet{planck/25:2015} compute the difference between the combined weighted \planck\ LFI and \wmap\ polarization P 
maps, defined as 
$[(Q_{\rm{LFI}} - Q_{\rm{\wmap}})^2 + (U_{\rm{LFI}} - U_{\rm{\wmap}})^2]^{1/2}$.  They also note large angular scale residuals at high latitudes, and attribute these
primarily to \wmap\ poorly measured modes.  Differences in the Galactic plane are assumed to arise from spectral index variations 
between frequencies that are not accounted for under the assumption of a fixed high Galactic latitude spectral index.

The attribution of the residual signature to \wmap\ poorly measured modes  carries consequences only in terms of visual presentation,
as it is not a true systematic residual.  When constructing and viewing maps without accounting for variance, as we have in 
Figure~\ref{fig:k30_qusys}, there can appear to be large angular scale
structure because of the enhanced noise.  
However, within the context of a low multipole likelihood, where full covariance matrices 
are included, poorly measured modes are correctly weighted in the analysis \citep{page/etal:2007}.

\begin{figure*}
\begin{center}
\includegraphics[height=5.0in]{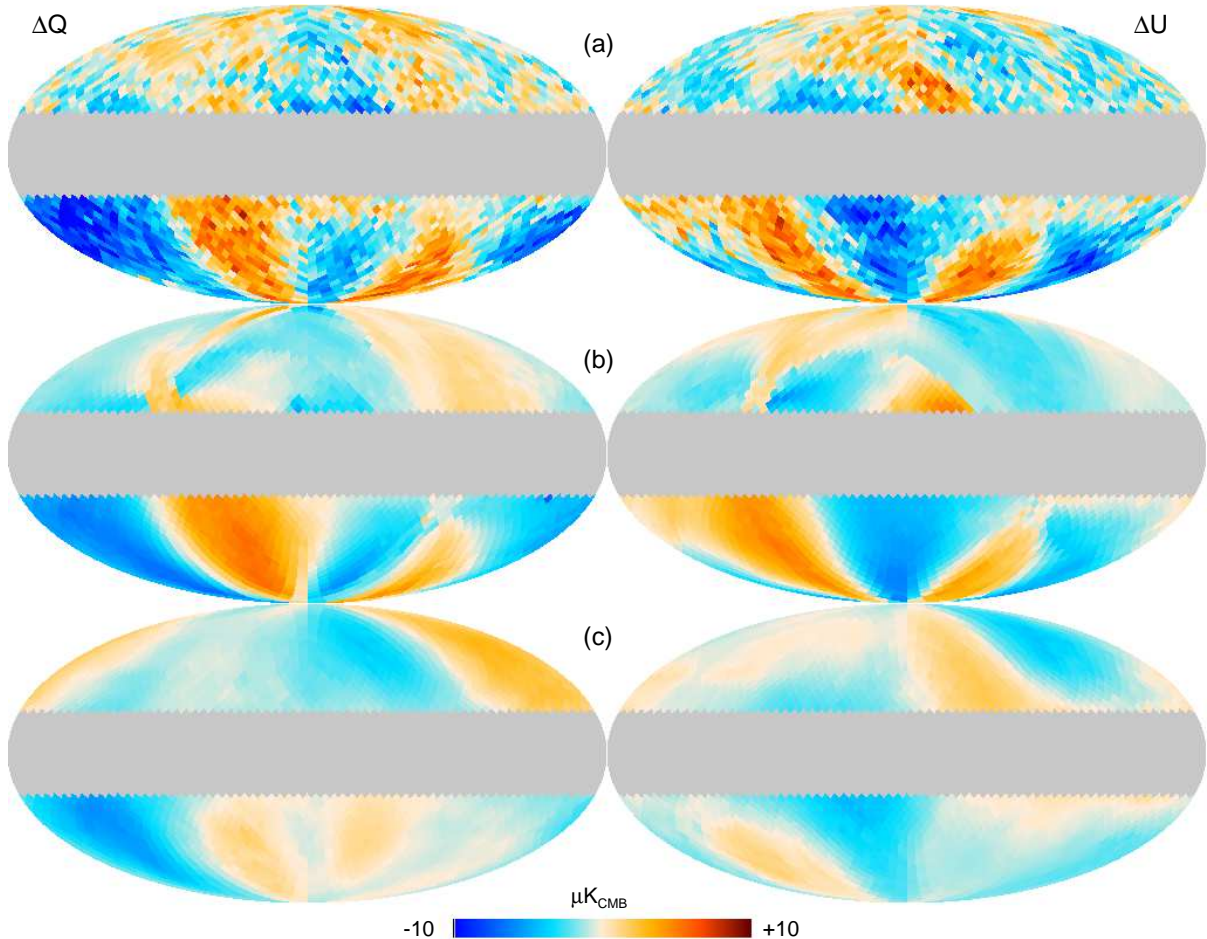}
\end{center} 
\vspace{-0.25cm}
\caption{ 
Based on morphological arguments, we conclude that \planck\ contributions to the large-scale [$\Delta$Q,$\Delta$U] signature 
in Figure~\ref{fig:k30_qusys} cannot be discounted and are likely the dominant source.
\textit{(a)} Same as Figure~\ref{fig:k30_qusys}b, i.e.  polarized LFI 30 GHz  minus \wmap\ K scaled to $\nu_{\rm{eff}}= 30$ GHz.
We fit
combinations of likely noise and calibration-related templates to this signature (see text), and present fits 
that attribute the morphology either wholly to
\planck-related causes or wholly to \wmap-related causes.
The grey band is a $\pm20^{\circ}$ cut about the Galactic plane, defining the fit mask.
\textit{(b)} An empirical fit to the $\Delta$Q and $\Delta$U pattern in  (a) using templates that assume the ßstructure 
is attributable to leakage of \planck\ intensity into Q and U (see text).
\textit{(c)} As with (b), but using templates characterizing \wmap\ poorly measured modes on large angular scales and
loss imbalance uncertainty.
\label{fig:k30_qusys_fits}
}
\end{figure*}

Isolating the root source of any residual signature that may be present in the \texttt{Commander} $S_{30}$ Q, U maps is complicated by the nature of its
production, as it relies on assumed foreground modeling dependencies and
is derived by fitting these models to a combined set of LFI and HFI polarization maps.
Therefore, we simplify the analysis
by computing the difference between two frequencies only: the full-mission LFI
30~GHz and \wmap\ K-band Q, U maps, both scaled to the effective frequency of 30 GHz using a $\nu^{-3.2}$ dependence: 
$\Delta [Q, U]= 0.85 [Q, U]_{LFI~30} - 0.39 [Q, U]_{\rm{\wmap\ K}}$, shown in Figure~\ref{fig:k30_qusys}b at \texttt{HEALPix} resolution $N_{\rm{side}}=16$.  
(All maps in Figure~\ref{fig:k30_qusys} are shown at a fiducial frequency of 30 GHz for ease of intercomparison).
Although ``$30 -$K'' (Figure~\ref{fig:k30_qusys}b) is slightly noisier in appearance than $``S_{30} - $K'' (Figure~\ref{fig:k30_qusys}a), the 
spatial correspondence between high-latitude features 
is quite similar between the two: the linear Pearson correlation coefficient is 0.9 for Galactic latitudes $|b| > 10^\circ$.
Since the large-scale signature is identified via a difference between LFI and \wmap\ data, its origin
may lie with either \planck\ or \wmap\ data, or both.  However, Figure~\ref{fig:k30_qusys}b illustrates that if there is a \planck\
contribution to this signature, then 30 GHz is involved, not only 44 GHz as suggested by \citet{planck/10:2015}.  

In Figure~\ref{fig:k30_qusys}c, we also show the difference between $S_{30}$ and LFI 30 GHz,
or equivalently between $``S_{30} - $K'' and $``30 - $K''.  This last figure illustrates the presence of large angular scale  
structure associated with \planck\ analysis products only, although the difference is between data and a model that fits that data.
However, \citet{planck/10:2015} illustrate examples of the model's ability to fit actual sky signal while isolating
residuals associated with calibration artifacts, as in the destriper signatures in temperature data.  It is likely then that the
``$S_{30}-30$'' residual is associated with calibration artifacts in \planck\ polarization data \citep[see also][]{wehus/eriksen:2017}.
This \planck-related large angular scale structure shares common morphological features with $``S_{30} - $K'' and $``30 - $K'', although not
necessarily of the same sign.
As there is no publicly available \texttt{Commander}-like foreground polarization model derived using only \wmap\ data,  an  
equivalent $``S_{\rm{K}} -$K'' map is not shown.  The \wmap\  MCMC foreground polarization models presented by \citet{bennett/etal:2013}
cannot serve as estimators of potential K-band residual structure, since certain model
parameters are specifically fixed to values computed directly from K-band data.  The morphology and amplitude of K-band poorly measured 
modes (a noise property and not a systematic error) are discussed in the next section.
We note that the differences shown in Figure~\ref{fig:k30_qusys} are not particularly sensitive to the choice of spectral index for 
scaling maps to the fiducial frequency, that the magnitude of the residual is not consistent with spectral index variations over the sky,
and that the exclusion of LFI surveys 2 and 4 does not appreciably alter the morphology shown.

\subsection{Morphological template fits}

We attempted to determine if plausible mechanisms associated with calibration uncertainties and/or noise properties could 
produce the morphology of the ``30 - K'' residuals shown in Figure~\ref{fig:k30_qusys}b.  
Specifically, we constructed Q, U template maps whose origin could be associated with a
physically-motivated mechanism arising either solely from \planck\ LFI 30~GHz, or solely from \wmap\ K-band.  
For \planck\ LFI, the documented source of largest uncertainty lies with the correction for 
leakage of intensity into polarization \citep{planck/02:2015}.  We constructed a set of templates to reflect possible differences
between the ``true'' leakage correction and that actually applied.   We used the 30~GHz full mission temperature map and transformed it to 
Q, U leakage maps via cosine and sine of the geometrical angle computed from the IQ and IU full mission covariance maps, and also 
performed a similar
computation of far-sidelobe leakage templates using the temperature maps from Figure 24 of \citet{planck/03:2013}.  The templates
are arbitrarily normalized since they are designed to approximate morphology only and do not reflect leakage efficiency.
The delivered 30~GHz BPM full mission Q, U correction maps were also included in the LFI template set.  For \wmap\ K-band,
we include Q, U templates for the poorly measured modes, and the two forms of loss imbalance correction maps  \citep{jarosik/etal:2007}
included in the \wmap\ final product delivery. 
A consideration in choosing templates was to limit the basis set to the most likely candidates, and include nothing further, since
a sufficiently large number of templates and degrees of freedom can reproduce almost any morphology without
regard to actual physical cause. 
We empirically fit the large angular scale high latitude ``30 - K'' $\Delta$Q, $\Delta$U signatures 
using an unweighted linear combination of spatial templates described above. Fits were performed to $\Delta$Q, $\Delta$U data outside of 
$|b| \geq 20^{\circ}$, in order to exclude Galactic plane regions most affected by spectral index variations and 
BPM correction uncertainties, where the template would be least accurate. 
$\Delta$Q and $\Delta$U were fit separately to the \wmap\ and LFI template basis sets.
Figure~\ref{fig:k30_qusys_fits}b shows the resulting fit using the LFI template basis set, while Figure~\ref{fig:k30_qusys_fits}c
shows that using the \wmap\ template basis set. These may be compared visually against the input systematic maps in the top
row of the figure.
Neither of the combinations exactly reproduce the input [$\Delta$Q, $\Delta$U] signature: the linear correlation coefficient between 
[$\Delta$Q, $\Delta$U] and the two empirical fits is 0.76 for the LFI basis set, and 0.46 for the \wmap\ basis set. 
The somewhat stronger morphological agreement between the input signature and the LFI fit, coupled with the apparent presence of 
large angular scale structure isolated to \planck\ alone, leads us to suspect a dominant contribution from the LFI 30~GHz map.

Amplitude arguments also favor a stronger association with the LFI 30~GHz map.   When the empirical templates in Figure~\ref{fig:k30_qusys_fits}b and c are peak-normalized
and used as basis functions in a linear fit to the ``30 - K'' $\Delta$Q, $\Delta$U signatures, we derive
an LFI template coefficient of $5~\mu$K, but only  $0.7~\mu$K for the \wmap\ template.  The \wmap\ template coefficient is within expectations
for poorly measured modes described by the K-band covariance matrix and extrapolated to 30~GHz.

\begin{figure*}[ht]f
\begin{center}
\includegraphics[width=6.5in]{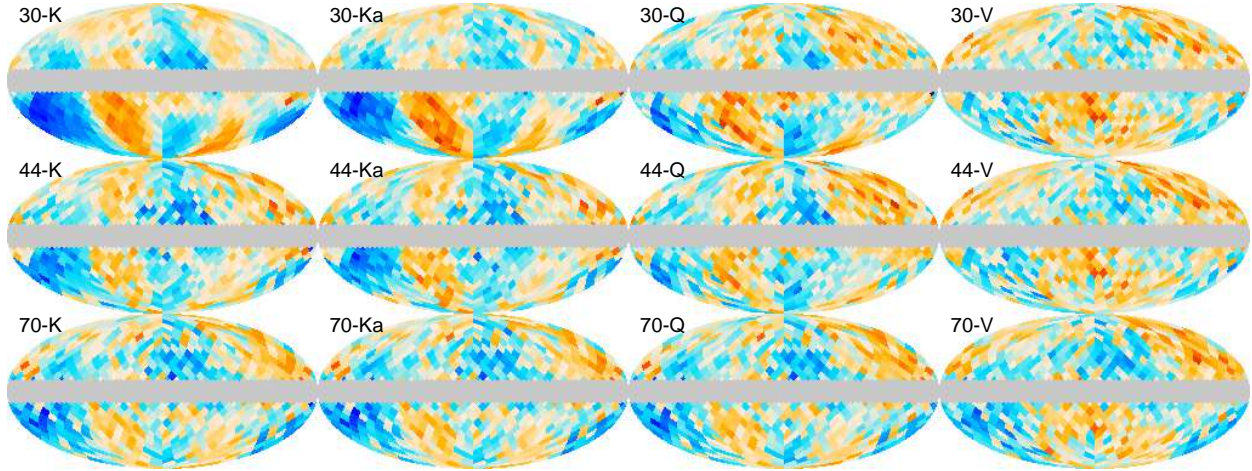}
\end{center} 
\vspace{-0.25cm}
\caption{Stokes Q foreground-nulled band-pair maps for \wmap-LFI frequency pairs. The grey band
excludes Galactic latitudes $|b| < 10^{\circ}$.  
Synchrotron foreground has been nulled using a $\nu^{-3.2}$ spectral dependence,
and dust emission removed from V-band and 70~GHz using the \planck\ \texttt{Commander} foreground model.
The resultant residuals reflect instrument noise,  cosmological signal and imperfections in foreground removal  that are both
primarily subdominant to the noise, and a large-scale signature with strong features
of order $5-10 \mu$K in the 30-K combination. 
Each band-pair image is autoscaled using an outlier-resistant determination of extrema, so that
the morphology in each image is readily apparent.
The residual large-scale signature is most clearly visible in the 30-K combination, but visually persists
in other frequency combinations. 
Taken in context with Figure~\ref{fig:u_bandpairs}, we infer a common-mode systematic across frequencies.
\label{fig:q_bandpairs}
}
\end{figure*}

\begin{figure*}[ht]
\begin{center}
\includegraphics[width=6.5in]{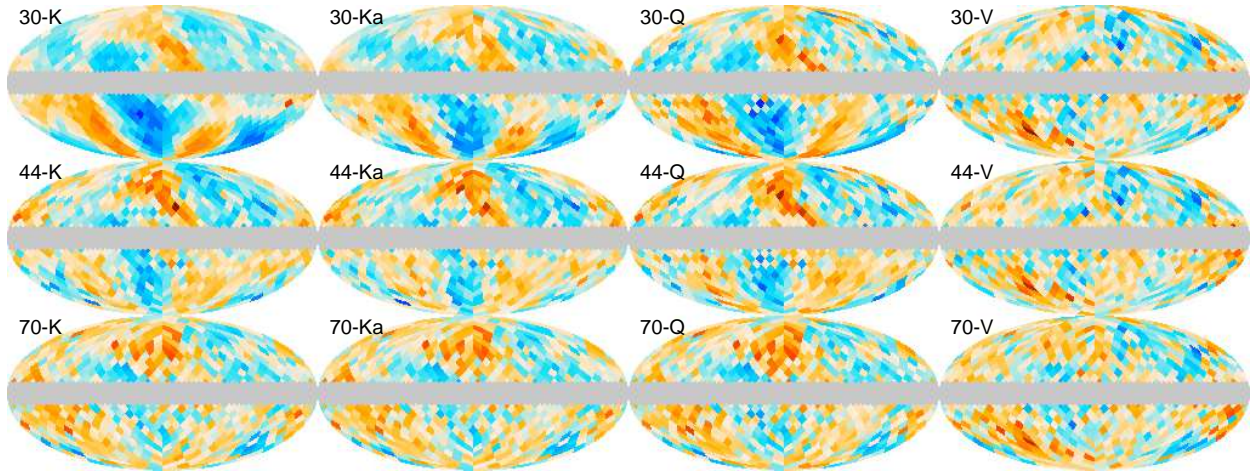}
\end{center} 
\caption{Same as for Figure~\ref{fig:q_bandpairs}, but for the corresponding Stokes U nulled 
band-pair maps for \wmap-LFI pairs.  See Figure~\ref{fig:q_bandpairs} caption for further explanation.
\label{fig:u_bandpairs}
}
\vspace{0.25cm}
\end{figure*}

\begin{figure}[ht]
\begin{center}
\includegraphics[height=3.5in]{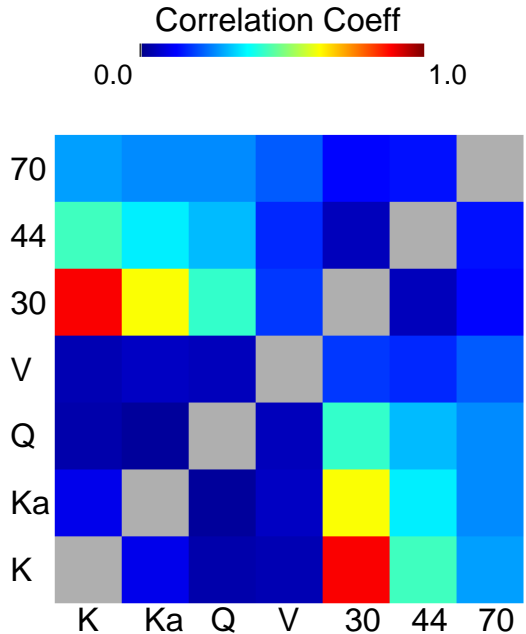}
\end{center} 
\caption{The absolute value of the linear correlation coefficient between the ``$S_{30}-K$''  
map shown in Figure~\ref{fig:k30_qusys}a 
and each of the LFI and \wmap\ [Q, U] band-pair difference maps shown, for example, in
Figures~\ref{fig:q_bandpairs} and \ref{fig:u_bandpairs}.  By construction, the matrix is symmetric. 
The correlation coefficients for \wmap-LFI pair differences are shown in the upper left quadrant and 
duplicated in the lower right.
The lower left quadrant shows the correlation for \wmap-internal combinations, and the upper right
for LFI-internal combinations.  Only \wmap-LFI band-pair combinations show significant correlation
with the systematic.  We deduce that the systematic signature has a frequency dependence that matches
the nulled foreground spectrum in the band-pair maps (see text).
\label{fig:qu_bandpairs_lincorr}
}
\end{figure}

However, if the signature is limited to the LFI 30~GHz band, we would expect greater disagreement between values for 
$\tau$ derived from the \wmap-only and LFI-only fits in rows 2 and 4 of Table~\ref{tab:tau_results_tab}. A 5-10 $\mu$K systematic 
in the 30~GHz synchrotron
template would scale down to roughly a 0.3 - 0.7 $\mu$K systematic at 70~GHz, which is sufficiently large compared to the predicted
polarized CMB signal to cause noticeable bias in recovered values of $\tau$.

\subsection{Spectral behavior}

We investigated to what extent the [$\Delta$Q, $\Delta$U] signature could be detected in \wmap\ and LFI frequency combinations other 
than 30 GHz and K-band. We examined both \wmap-only and LFI-only combinations as well as \wmap-LFI combinations involving \wmap\ K, Ka, Q and V bands, and
the LFI 30, 44 and 70 GHz bands. 
We formed ``band-pair'' maps from the weighted linear combination of two independent frequencies. 
Weights for each frequency were computed to null a
$\nu^{-3.2}$ brightness temperature spectrum, yet preserve the underlying CMB signal, such that $\rm{wt}(\nu_{1}) + \rm{wt}(\nu_{2}) = 1$
when applied to maps in thermodynamic temperature units.  
Because only one foreground spectral constraint can be applied with this weighting scheme, we removed the \texttt{Commander} polarized dust emission model
from all frequencies beforehand. The dust emission removal is of minor importance for all frequencies except V-band and 70~GHz.
Identical weights are applied to the Q and U maps of each band-pair.  Stokes Q \wmap-LFI pair differences are shown in 
Figure~\ref{fig:q_bandpairs}, and 
those for Stokes U in Figure~\ref{fig:u_bandpairs}.  Each individual band-pair image is autoscaled using an outlier-resistant minima 
and maxima determination scheme, so that neighboring images may be more easily compared visually despite noise variations.  
Features become less distinct 
for band-pair combinations that include at least one of the higher frequency maps, as their signal-to-noise is lower.
With the exception of the 70-V combination,
combinations including 70 or V are heavily weighted toward these frequencies and do not change substantially in appearance as the companion
frequency varies. Combinations that include \wmap\ V-band include the visual signature from poorly measured modes.  The ``30-K''
signature persists in \wmap-LFI combinations involving K, Ka, Q, 30~GHz and 44~GHz, with less certainty about 70~GHz and V.
We infer a systematic error that is common-mode across frequencies.

We linearly correlated the ``$S_{30}-K$'' [Q,U] templates against each of the  \wmap- and LFI-internal combinations and
the \wmap-LFI pair combination [Q,U] maps.  The correlation was performed outside Galactic latitude $|b| \geq 10^{\circ}$, and treated [Q,U] 
simultaneously in the solution.  We show the correlation coefficients as a symmetric matrix in Figure~\ref{fig:qu_bandpairs_lincorr}, 
and show the absolute value of the correlation coefficient, since sign is irrelevant in this context.  
Significant correlations are found only in the \wmap-LFI pair combinations.
The ``30-K'' signature does not 
correlate strongly in the LFI-internal combinations containing 30~GHz,  nor are there strong correlations in the \wmap-internal combinations 
that include K-band.  The simplest explanation is that the signature itself possesses a frequency
dependence that closely matches that of $\nu^{-3.2}$.  A systematic isolated to LFI Q, U maps with  
a frequency dependence that matches the nulled foreground spectrum would be undetectable in the LFI-internal
band-pair combinations, and only appear in \wmap-LFI  pair combinations.  Such a systematic would also evade LFI-internal null tests
based on in-band jack-knife tests.
The poorly measured modes associated with \wmap\ do not match a foreground spectrum signature, whereas such a spectrum is 
a natural consequence from leakage of strong Galactic temperature signal into LFI polarization maps.
The \wmap\ sky scan pattern probes a larger range of scan angle orientations than does \planck's scan pattern, so it is reasonable to expect that the leakage would be of far greater concern for \planck\ than for \wmap\ polarization data   \citep{planck/02:2015, page/etal:2007}.

\subsection{Effect of increased masking and multipole cuts}

We applied two common techniques that attempt to
mask or otherwise mitigate the systematic signature introduced when cleaning LFI data with a \wmap\ synchrotron template and vice-versa.
In the first method, we constructed a variety of masks seeking to define smaller areas of sky less affected by non-CMB signals.  
Our most restrictive mask excluded regions of the Q and U ``30-K'' signature with absolute amplitudes  $>2 ~\mu$K, in combination with the \planck\ 
R1.50 analysis mask, retaining only 11\% of the sky. Chains duplicating rows 7 and 8 of Table~\ref{tab:tau_results_tab} but using this new mask returned $\tau$ for \wmap\ cleaned with LFI~30~GHz of $0.24 \pm 0.03$, and $0.09 \pm 0.03$ for LFI~70~GHz cleaned with \wmap\ K-band.   
None of the masks we explored produced mean $\tau$ values similar to those found when using \wmap-internal or LFI-internal data and synchrotron template combinations.  The second technique excluded data in harmonic space but retained the original R1.50 pixel analysis mask.
Motivated by a statement in 
\citet{planck/10:2015}  that the ``S$_{30}-$K'' signature is  dominated by modes for multipoles 2 and 3,  we excluded these multipoles and re-ran chains
under the same foreground cleaning and masking conditions as specified in rows 7 and 8 of Table~\ref{tab:tau_results_tab}.   This method produced $\tau$ values less
than $1\sigma$ different from those derived using the full multipole range.

\subsection{Implications for $\tau$}

What are the consequences for $\tau$ 
with an LFI temperature to polarization leakage that matches the sky spectrum?
As long as LFI 30~GHz is used as the synchrotron template for LFI 70~GHz, and the morphology of the
systematic is frequency independent, then the recovered values for $\tau$ should be relatively unaffected if the template foreground
coefficients accurately reflect the foreground contribution.  
The relative agreement between the \wmap-internal and LFI-internal $\tau$ results
under the same masking conditions implies at least some consistency with this picture.
However, we cannot quantitatively assess the degree to which these conditional statements are met, and there may
be additional issues such as somewhat different spectral index dependencies between temperature and polarization.
Use of LFI 30~GHz as a template for cleaning non-LFI datasets 
such as HFI data would produce a biased result, assuming there is no equivalent signature in the other dataset.
The 2016 HFI $\tau$ analysis \citep{planck/intermediate/46:2016} makes use of LFI polarization data in several ways: as templates for residual
systematic evaluation, as foreground templates (30~GHz), and in EE cross-spectra (70x100~GHz, 70x143~GHz) for confirmation of results from
100x143~GHz.  Synchrotron emission is not a large contributor at 100 GHz, and even less at 143~GHz, so bias introduced by
use of the LFI 30~GHz template should be low.
We suggest a useful test of robustness of the HFI result would include substituting equivalent \wmap\ data for LFI data in the
analysis.

\section{Conclusions}

The reionization optical depth,  $\tau$, is currently the least well constrained of the six standard \lcdm\ parameters.
Large angular scale microwave polarization data provide the means for direct measurement of $\tau$, but Galactic
foreground emission must be removed with sub-$\mu$K accuracy to expose the underlying cosmological signal. 
Foreground removal using templates, a technique employed
by both \wmap\ and \planck, introduces uncertainties associated with accuracy of the templates,
including decoherence of the template morphology with frequency and systematics in the templates.
Instrumental noise and data systematics also contribute to uncertainties, 
requiring some tailoring of analysis choices to individual datasets.

In this paper, we have pursued combined analysis of \wmap\ and \planck\ LFI polarization data
as a means of cross-validating published $\tau$ values derived using publicly unavailable HFI 100 and 143~GHz 
Stokes Q and U maps.  In doing so, we have encountered a few issues:

\begin{enumerate}

\item Analysis mask selection techniques involving  $\chi^2$ minimization act to maximize bias from chance correlations of CMB
signal with foreground templates, noise and systematics.  The  importance of this bias term in the uncertainty budget 
depends on what additional sources of map-cleaning error are present and their relative dominance.
In general, there is no substitute for carefully assessing each specific situation.
We recommend a systematic approach to analysis mask selection through simulations.

\item There is an inconsistency between the 353~GHz Q and U thermal dust template maps internal to the \planck\ low multipole 
likelihood code, and those available as delivered maps from the \Planck\  Legacy Archive.  The mismatch
produces a $1\sigma$ shift in $\tau$ derived from the LFI 70 GHz data.

\item An unexpectedly high value for $\tau$ is derived when \wmap\ K-band Stokes Q and U maps are 
used to clean synchrotron emission from polarized \planck\ LFI 70~GHz data, and a severely compromised $\tau$ value results when
LFI~30~GHz Q and U maps are used to clean synchrotron from the combined \wmap\  Ka, Q and V bands.

\end{enumerate}

Template systematics are the root cause of the synchrotron cleaning issue.
Differences between combined LFI polarization bands and their \wmap\ frequency counterparts 
have revealed residual structure on large angular scales  \citep{planck/10:2015, planck/25:2015}.
The \Planck\ Collaboration concluded that this residual pattern primarily
arises from \wmap\ poorly measured modes, with any \planck-related contribution limited to the LFI 44 GHz band.
We reject both of these assertions and show evidence that the signature is a systematic internal to LFI and 
roughly common-mode to all LFI bands.
The spectral behavior of the systematic scales like
the intensity signal and is thus likely to arise from
temperature-to-polarization leakage in the LFI channels. 
In an LFI-only analysis this systematic is largely nulled by the foreground removal procedure,
but attempting to
combine \wmap\ with \planck\ data before foreground removal creates an
inconsistency that greatly compromises the use of both data sets together.
We were not able to mitigate the effects of the systematic either through additional masking or removal of multipoles 2 and 3.

\citet{planck/11:2015} derive $\tau = 0.071 \pm 0.012$ from a combined \wmap+LFI analysis (see Table~\ref{tab:tau_pub_tab}).
This result is based upon a weighted coaddition of cleaned maps consisting of \wmap\ data cleaned with K-band and 353~GHz over 75\% of the sky,
and LFI 70 GHz data cleaned with LFI 30~GHz and 353~GHz over 46\% of the sky.  Although such a combination produces a
result for $\tau$ that is not unreasonable, we cannot assess the associated systematic uncertainty accurately.
In addition to the mix of analysis masks, the LFI temperature to polarization leakage efficiency may not be constant with frequency, 
and production of a correction template would be circular since at present it would be based on \wmap\ and LFI data.
As a comparison standard for other $\tau$ determinations, we believe this particular result is compromised, but
should be re-examined when an updated LFI calibration becomes available.

Unfortunately, the levels of systematics, foregrounds, and the quality of the foreground templates are not often well known. Robustness can be 
estimated by fitting a range of CMB maps with a range of templates under different masking assumptions. 
We note the spread in results for $\tau$ both in Table~\ref{tab:tau_pub_tab} and Table~\ref{tab:tau_results_tab} (excluding rows 7 and 8), and
comment that the quoted uncertainties for $\tau$ are most likely underestimated because systematic
uncertainties are not included.  However, tension is not high 
between the quoted values because of the relatively large uncertainties and all are 
consistent with a polarization-derived value of $\tau = 0.07 \pm 0.02$,  used as a prior in some earlier studies \citep{planck/11:2015, 
addison/etal:2016, aylor/etal:2017}.  Values for $\tau$ derived without the use of polarization data are also consistent with this range.
\citet{planck/intermediate/46:2016} derived $\tau = 0.067 \pm 0.016$ for \planck\ TT+lens+BAO
(see Table~\ref{tab:tau_pub_tab}).  Adopting the analysis methods and BAO dataset selection of \citet{addison/etal:2017},
we derive a similar value, $\tau = 0.066 \pm 0.020$,  for the combination of  \wmap\ 9-year TT,  \planck\ 2015 weak lensing, and BAO.

There is considerably more work to be done in terms of characterizing $\tau$ and choosing the optimal
analysis masks and foreground removal given instrumental properties.  Future datasets 
should also assist with the analysis.  These include the final \planck\ data release, additional
weak lensing constraining power from ground-based surveys such as ACTPol and SPTpol, and
high signal-to-noise polarization observations from e.g. the Cosmic Large Angular Scale Survey (CLASS), which is projected
to recover $\tau$ to nearly cosmic variance uncertainties and enable a $4\sigma$ detection of the
sum of the neutrino masses \citep{watts/etal:2018}.

\vspace*{0.15in}
%\acknowledgements
This research was supported in part by NASA grants NNX16AF28G, NNX17AF34G and by the Canadian Institute for Advanced Research (CIFAR).  
We acknowledge use of the \texttt{HEALPix} \citep{gorski/etal:2005} and \texttt{CAMB} \citep{lewis/challinor/lasenby:2000} packages.  
This research has made use of NASA's Astrophysics Data System Bibliographic Services.  We acknowledge the use of the 
Legacy Archive for Microwave Background Data Analysis (LAMBDA), part of the High Energy Astrophysics Science Archive Center (HEASARC). 
HEASARC/LAMBDA is a service of the Astrophysics Science Division at the NASA Goddard Space Flight Center.  
We also acknowledge use of the \Planck\ Legacy Archive. \Planck\ is an ESA science mission with instruments and contributions 
directly funded by ESA Member States, NASA, and Canada.

\appendix
\section{353~GHz Dust Template}
\label{sec:dust_template}

Full-sky low-resolution ($N_{\rm{side}}=16$) 353~GHz Q and U templates are not available from the \Planck\ Legacy Archive.
Section 2.3 of \citet{planck/11:2015} describes the dust templates used in cleaning the LFI 70~GHz Stokes Q and U maps
as bandpass-mismatch (BPM) corrected 353 GHz Q, U maps  degraded to $N_{\rm{side}}=16$ using inverse noise weighting. 
There are however two publicly available options for 353 GHz bandpass mismatch corrections.  
The first of these
is the default option, or ``ground/dust leakage correction'', which is computed 
through application of the map-making pipeline to model dust and CO emission components, in conjunction with  transmission coefficients derived by
integration of the component spectral response over the  ground-based measurements of the bandpass \citep[see Section 7.3.1] {planck/08:2015}.
A second option, or ``global leakage correction'', was still in the experimental phase in 2015.
As described by \citet{planck/08:2015}, this algorithm attempts to simultaneously incorporate bandpass, monopole and calibration leakage 
into a single correction map set.  Maps of both correction types are shown for the region outside the \planck\ low multipole polarization 
analysis mask (R1.50) in Figure~\ref{fig:bpm353}.
The inclusion of a CMB dipole signature in the ``global'' correction is one feature that distinguishes it from the default ground/dust 
leakage sky pattern. 
The magnitude of both correction options at 353 GHz is of order $10~\mu$K at high Galactic latitudes.
Extrapolation of this signal to 70~GHz using the \planck\ 2015 dust template coefficient of 0.0077 mitigates the correction contribution 
to a level roughly 3 times smaller than the expected rms of the CMB signal at $N_{\rm{side}}=16$ (0.077 vs. 0.25 $\mu$K), and 
\citet{planck/02:2015} note that changes to the dust
template of this order should not have a large effect on the recovered value of $\tau$ from the 70 GHz data.

\begin{figure*}
\begin{center}
\includegraphics[width=6.0in]{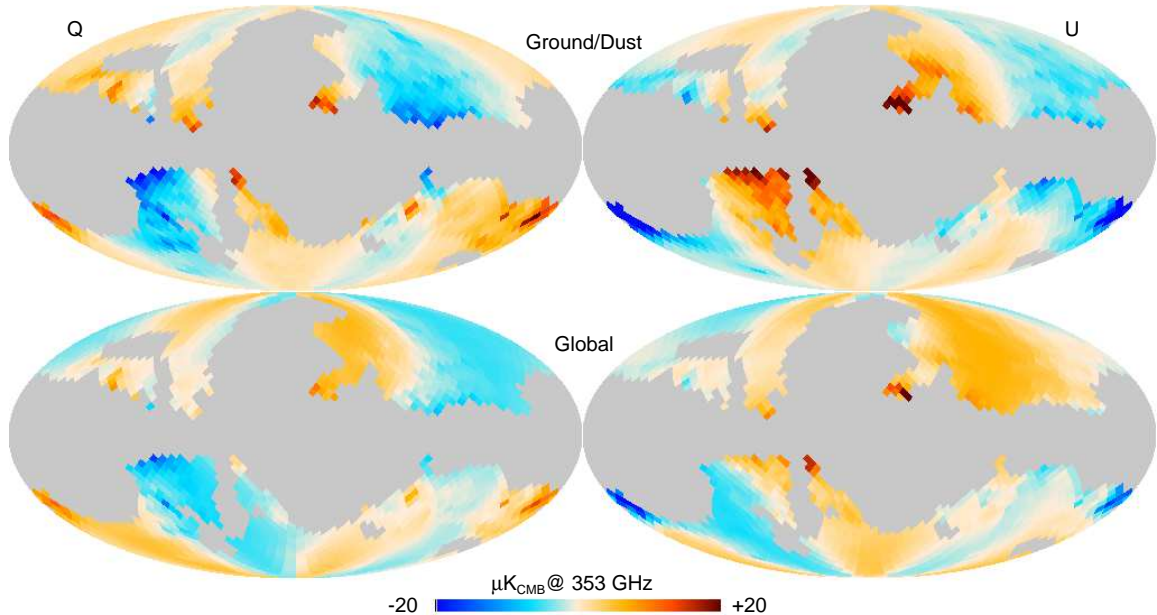}
\end{center} 
\vspace{-0.25cm}
\caption{Morphology and magnitude of delivered \planck\ 2015 353 GHz dust template leakage systematic corrections for Stokes Q and U.
 \textit{Top}: The ground/dust leakage corrections, which are by default used to correct the delivered \planck\ 353~GHz full mission map.
 \textit{Bottom}: The global leakage correction, which attempts to include both calibration and bandpass mismatch leakage effects,
 which was made available but not applied to the 2015 products.  The global correction has a morphology that is distinguishable from that of the 
 ground/dust correction.
\label{fig:bpm353}
}
\end{figure*}

\begin{figure*}
\begin{center}
\includegraphics[width=6.0in]{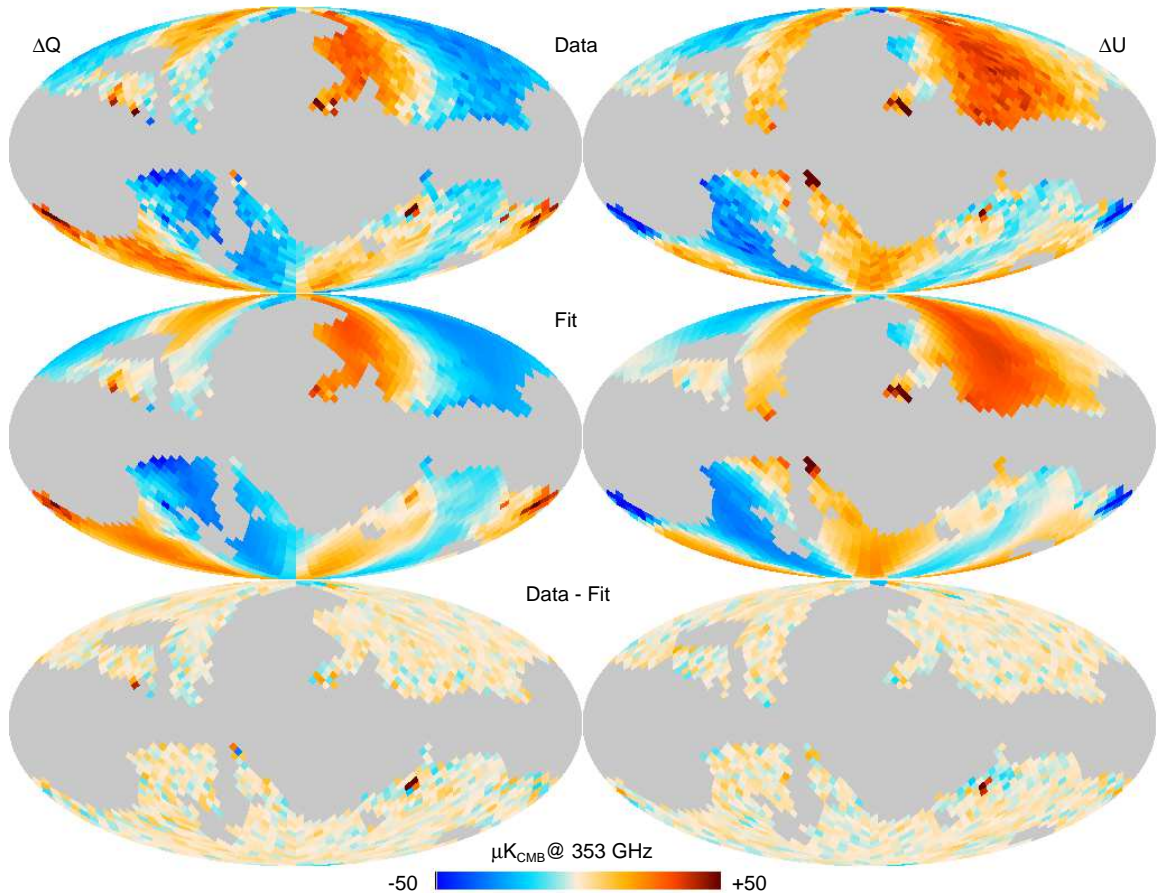}
\end{center} 
\vspace{-0.25cm}
\caption{We attempt to verify which BPM correction was applied to the 353 GHz dust template used in the low multipole \texttt{plc} likelihood.
 \textit{Top}: the difference between the 353 GHz dust template derived from the \texttt{plc} likelihood (see equation A1) and the
 delivered 353 GHz map degraded to $N_{\rm{side}}=16$. This difference represents the difference between the default ground/dust 
 leakage correction and the leakage correction used for the \texttt{plc} 353 GHz dust template (see text for detail).
 \textit{Middle}: a fit to the top row residuals based on a linear combination of the ground/dust leakage template
 and the global leakage template.  The fit is dominated by the global leakage correction term, but with a
 5 times larger amplitude.
 \textit{Bottom}: residuals of the top row minus the fit in the middle row. 
 The apparent magnitude of the correction applied to the dust template places its contribution at 70~GHz near CMB signal levels, 
 and may affect the determination of $\tau$.
\label{fig:corr_surprise}
}
\end{figure*}

We attempted to verify which version of the leakage correction was applied to the 353 GHz dust template used to produce the cleaned 70 GHz QU
maps delivered in the \planck\ likelihood.  To do this, we solved for the masked 353 GHz Q, U template contribution, $m_{353}$, 
based on the cleaned map
contained within the likelihood, $m_{cln}$, and all other known, delivered, components as related by equation 7 of \citet{planck/11:2015}
\begin{equation}
m_{cln} = (1-\alpha_s-\alpha_d)^{-1} (m_{70} - \alpha_s m_{30} - \alpha_d m_{353})
\end{equation}
where $m_{70}$ is the observed 70 GHz [Q, U] after BPM correction, $m_{30}$ is observed 30 GHz [Q, U] after BPM correction and acts as the
synchrotron template, and $\alpha_s, \alpha_d$ are the synchrotron and dust template coefficients, respectively, with values 0.063 and 0.0077
as specified in \citet{planck/11:2015}.

The difference between the default full mission 353 GHz Q and U maps, which have been corrected for BPM using the ``ground/dust leakage'' 
maps, and
the 353 GHz template we derive from the \texttt{plc} cleaned 70 GHz map is shown in the top row of Figure~\ref{fig:corr_surprise}.  
This difference should be 
approximately zero everywhere if the \texttt{plc} dust template were equivalent to the delivered versions of the corrected full mission 353 GHz QU maps, assuming the baseline uncorrected observed 353~GHz maps are in common:
$ \Delta m_{353} = m_{353}^{\rm{plc}} - m_{353}^{\rm{default}} =  {BPM}^{\rm{default}} - {BPM}^{\rm{plc}}$. We do expect
small high frequency differences in that we did not degrade the 353 GHz map using inverse noise weighting as did \planck, but rather flat 
weighting after smoothing by a $2^{\circ}$ FWHM Gaussian.
However, the difference predominantly shows 
large scale coherent signatures of order $30~\mu$K, whose morphology resembles that of the global correction templates shown in Figure~\ref{fig:bpm353}, 
but with a larger amplitude, and cannot be attributed to differences in the map downgrading algorithm.
We can fit the difference shown in the top row of this figure with a linear combination of both the ground/dust and global leakage templates:
$\Delta m_{353} = A \times BPM^{\rm{ground}} + B \times BPM^{\rm{global}}$.  An unweighted fit returns coefficients $A = -1.00$ and $B = 5.36$.
The fit result implies that the 353 GHz dust template used to clean the 70 GHz map contained within the \texttt{plc} likelihood was corrected with
neither of the delivered leakage correction maps, but rather a correction with global leakage morphology and a 5 times greater amplitude.
This is a puzzling and unanticipated result.  A $30~\mu$K leakage correction at 353 GHz extrapolates roughly to
$0.23~\mu$K at 70~GHz, which is of comparable amplitude to the expected $N_{\rm{side}}=16$ rms CMB signal for $\tau = 0.05$.

Ongoing updates to the HFI calibration pipeline discussed in \citet{planck/intermediate/46:2016} indicate that the global leakage algorithm may 
ultimately provide a more realistic assessment of the BPM correction.  However, the delivered 2015 global leakage template is described
as only preliminary, and, as discussed here, its amplitude is uncertain.
We found no reason to favor one template over the other and thus included both \texttt{plc} and default options for the 353 GHz polarized dust 
emission template in our fits for $\tau$; see Table~\ref{tab:tau_results_tab} and Section~3.

\end{document}